\documentclass{emulateapj}
\usepackage{apjfonts}

\begin{document}

\title{Radial Velocity Detectability of Low Mass Extrasolar Planets in
Close Orbits}

\author{Raman Narayan, Andrew Cumming\altaffilmark{1,2}, and
D.~N.~C.~Lin} 
\affil{Department of Astronomy and Astrophysics,
University of California, Santa Cruz, CA 95064}
\altaffiltext{1}{Hubble Fellow} 
\altaffiltext{2}{Current address: Department of Physics,
McGill University, 3600 rue University, Montreal, QC, Canada H3A 2T8.}
 
\begin{abstract}
Detection of Jupiter mass companions to nearby solar type stars with
precise radial velocity measurements is now routine, and Doppler
surveys are moving towards lower velocity amplitudes. The detection of
several Neptune-mass planets with orbital periods less than a week has
been reported. The drive toward the search for close-in Earth-mass
planets is on the agenda.  Successful detection or meaningful upper
limits will place important constraints on the process of planet
formation. In this paper, we quantify the statistics of detection of
low-mass planets in-close orbits, showing how the detection threshold
depends on the number and timing of the observations. In particular,
we consider the case of a low-mass planet close to but not on the 2:1
mean motion resonance with a hot jupiter. This scenario is a likely
product of the core-accretion hypothesis for planet formation coupled
with migration of jupiters in the protoplanetary disk. It is also
advantageous for detection because the orbital period is
well-constrained. We show that the minimum detectable mass is $\approx 4\
M_\oplus\ (N/20)^{-1/2} (\sigma/{\rm m\ s^{-1}})(P/{\rm
d})^{1/3}(M_\star/M_\odot)^{2/3}$ for $N \ge 20$, where $N$ is
the number of observations, $P$ the orbital period, $\sigma$ the
quadrature sum of Doppler velocity measurement errors and stellar
jitter, and $M_\star$ the stellar mass. Detection of few Earth mass
rocky cores will require $\sim 1\ \mathrm{m\ s^{-1}}$ velocity
precision, and most important, a better understanding of stellar
radial velocity ``jitter''.
\end{abstract}


\section{Introduction}

Over 100 Jupiter-mass planets have been discovered around nearby stars with precise radial velocity measurements (see Marcy et al.~2003 for a review). As time goes by, the range of periods and amplitudes accessible to these surveys increases, moving to longer periods and lower amplitudes (e.g.~Carter et al.~2003; Fischer et al.~2003; Jones et al.~2003). Recently, the detection of Neptune-mass planets with periods less than ten days has been reported (Butler et al.~2004; McArthur et al.~2004; Santos et al.~2004). In this paper, we quantify the detectability of low mass planets in close orbits, and discuss the prospects for detecting earth mass objects. We first discuss possible origins of such planets and the importance of detecting them.

\subsection{The origin and importance of low mass rocky cores in close orbits}

In the conventional planet formation scenario (Safronov 1969; Pollack
et al.~1996), heavy elements coagulate into terrestrial-planet-like
cores prior to the formation of Jupiter-mass planets by subsequent
accretion of gas. The critical mass which segregates these two
populations is $M_c \sim$ a few $M_\oplus$, determined by the
requirement of sufficiently high cooling efficiency in the gas
envelope (Stevenson 1982; Bodenheimer \& Pollack 1986; Ikoma et
al.~2000). In a minimum mass nebula, both type of planets are thought
to form at distances from their host stars comparable to those of
planets in the solar system. 

As a consequence of their tidal interaction, gas giant planets may
undergo orbital migration (Goldreich \& Tremaine 1980; Lin \&
Papaloizou 1986) which is finally halted either by their interaction
with their host stars or through the local or global depletion of
disks (Lin et al.~1996; Trilling et al.~2002; Armitage et al.~2002;
Ida \& Lin 2004). In this process, many gas giants are expected to be
disrupted, transferring their mass onto the host star (Trilling et
al.~2002), perhaps driven by tidal inflation (Gu, Lin, \& Bodenheimer
2003). Ida \& Lin (2004) estimate that 90--95\% of planets that
migrate to $a\lesssim 0.05\ {\rm AU}$ must perish. Recent models of
evaporation of material from hot jupiters (Lammer et al.~2003; Baraffe
et al.~2004; Lecavalier des Etangs et al.~2004) find that gas giants
may undergo significant mass loss, and perhaps be completely
evaporated, during their lifetimes, although the efficiency and rates
of evaporation remain uncertain (Yelle 2004). This inference raises
the possibility of a class of remnant ``hot Neptunes'' or rocky cores
in close orbits\footnote{In this paper, we use the adjective ``hot'' to describe a closely-orbiting planet (orbital period $\lesssim 10$ days).}.

An alternative way to produce low-mass planets in close orbits is by
resonant capture during migration of a hot jupiter. Along their
migration paths, gas giant planets capture other planets onto their
mean motion resonances as in the case of GJ 876 (Lee \& Peale 2002;
Nelson \& Papaloizou 2002; Kley et al.~2004).  Mandell \& Sigurdsson
(2003) also consider the survivability of earth mass objects in the
wake of Jupiter migration.  A natural implication of the
core-accretion followed by migration scenario for the origin of hot
jupiters is that they may harbor earth-mass planets (``hot earths'') orbiting close to
their mean motion resonances (Aarseth \& Lin 2004, in preparation). In the
gravitational instability scenario, however, both ice giants and
terrestrial planets are assumed to emerge long after the formation of
gas giant planets (Boss, Wetherill, \& Haghighipour 2002).

Due to the effect of tidal circularization, the semimajor axis of the
hot earth is expected to be slightly interior to the mean motion
resonance, and its survival is ensured by the relativistic precession
induced by the gravitational potential of the host star (Novak et
al.~2003, Mardling \& Lin 2004).  After halting its inward migration,
a hot jupiter may move outwards due to either its tidally induced mass
loss (Trilling et al.~2002; Gu et al.~2003), or tidal interaction with
a rapidly spinning host star (Stassun et al.~1999; Dobbs-Dixon et
al.~2004). In both cases, the periods of the close-in earth and hot
jupiter may not be close to commensurability.

The presence of earth-mass planets close to the external mean motion
resonance of a hot jupiter is also possible. In the wake of a
migrating gas giant, the prototerrestrial planets in the disk external
to its orbit may also undergo type I inward migration (Ward 1986) as a
consequence of their own tidal interaction with the disk gas
(Goldreich \& Tremaine 1980).  By inducing the formation of a gap near
its orbit, a migrating jupiter provides a tidal barrier which prevents
any migrating terrestrial planets from passing it.  But this
possibility is uncertain since type I migration is strongly affected
by the poorly known turbulence near the coorbital region (Koller et
al.~2003; Nelson \& Papaloizou 2004). If this process is efficient, it
could lead to the formation of an isolated hot earth (Ward 1997;
Bodenheimer et al.~2000). Even if type I migration is inefficient, the
presence of a hot jupiter can induce planetesimals to accumulate and
form a hot earth just beyond the outer edge of the gap (Bryden et
al.~2000).  In the presence of additional long-period gas giant
planets, a sweeping resonance may also lead to the inward migration
and accumulation of terrestrial planets just exterior to the orbits of
hot Jupiters (Lin, Nagasawa, \& Thommes 2004).

Based on the core-accretion scenario for planet formation, it is
already tempting to infer the existence of terrestrial planets in
extrasolar systems from the detection of jupiter-mass gas giants (Ida
\& Lin 2004). However, the actual detection of a terrestrial mass
object would be the first confirmation of rocky cores in an extrasolar
planetary system (see however Kuchner 2003). The association of a hot
earth with the interior mean motion resonance of a hot jupiter is
particularly interesting because it would indicate that cores are
formed prior to the emergence and migration of gas giant planets,
providing strong support for the conventional core-accretion scenario
for planet formation, and for the ubiquity of terrestrial planets in
extrasolar planetary systems.  The presence of a hot earth exterior to
the orbit of a hot jupiter would support the notion of a tidal barrier
which has the potential to enhance the formation probability of
multiple planet systems. Finally, the discovery of any isolated hot
earth would support the concept of type I migration.

\subsection{Outline of this paper}

In this paper, we discuss the detectability of such low mass
companions in radial velocity surveys. In particular, we concentrate
on the scenario in which a terrestrial planet was captured into the
mean motion resonance of a migrating jupiter-mass planet, resulting in
the formation of a ``hot earth'' which is near the interior mean
motion resonances of a ``hot jupiter''. With the limitations of radial
velocity surveys, the detection of a hot earth external to the orbit
of a hot jupiter is much more difficult. We also calculate the
detection thresholds for isolated low mass planets.

Being offset from the mean motion resonance of any nearby hot Jupiters, the expected contribution to the radial velocity amplitude $K$ from a closely orbiting several Earth-mass object is 
\begin{equation}\label{eq:K}
K=6.4\ \mathrm{m/s}\ \left({M_P\sin i\over 10\ 
M_\earth}\right)\left({P\over\mathrm{days}}\right)^{-1/3}
\left({M_\star\over M_\odot}\right)^{-2/3},
\end{equation}
where $P$ is the orbital period, $M_P$ is the mass of the planet, and $M_\star$ is the mass of the star. This is comparable to the expected velocity variability from measurement errors, and intrinsic stellar ``jitter'' (Saar, Butler, \& Marcy 1998). The orbital frequency is well-predicted for a hot earth close to a mean motion resonance with a hot jupiter, and so we expect the detectability to be somewhat better than a blind frequency search. Here we calculate the detectability and discuss the number of observations, and observing strategy needed for detection.

\begin{figure}
\plotone{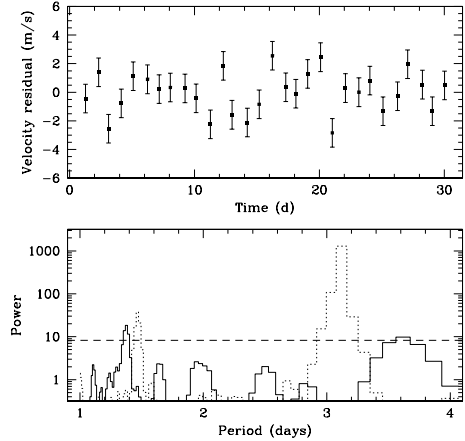}
\caption{Simulated velocity residuals (top panel) after subtraction of
a $0.5\ M_J$ mass planet with $P=3.1\ \mathrm{days}$, $K=70\ {\rm
m/s}$. We simulate one observation per night for 30 nights, with
measurement error of $1\ \mathrm{m\ s^{-1}}$. Periodogram (lower
panel) evaluated for periods from 1 to 4 days before (dotted line) and
after (solid line) subtraction. The strong feature present at $\approx
1.5$ days before subtraction is due to aliasing of the 3.1 day
signal. Subtracting reveals the presence of a $3\ M_\oplus$ companion
at $P=1.4$ days. The dashed line shows the detection threshold. 
\label{fig:periodogram}}
\end{figure}

\section{Detectability of Close Companions}

\subsection{Search Technique}

We simulate a set of observational data for a system of planets which
are on circular orbits.  We search for periodicities using the
Lomb-Scargle periodogram (Lomb 1976; Scargle 1982). For a set of
$N$ observation times $\left\{t_j\right\}$, velocities
$\left\{v_j\right\}$, and measurement errors $\left\{\sigma_j\right\}$, 
and a trial orbital frequency $\omega=2\pi/P$, we fit the function
\begin{equation}\label{eq:model}
f_j=A\cos\omega t_j+B\sin\omega t_j+C
\end{equation}
to the velocity data by minimising $\chi^2_\nu=(1/\nu)\sum_j
(v_j-f_j)^2/\sigma_j^2$. The number of degrees of freedom is $\nu=N-3$
since there are 3 parameters in the model. We measure the goodness of
fit as a function of trial frequency $\omega$ using the periodogram
power $z$, defined as
\begin{equation}\label{eq:z}
z(\omega)={\Delta \chi^2/2\over \chi^2_\nu},
\end{equation}
where $\Delta\chi^2=(N-1)\chi_{N-1}^2-\nu\chi^2_\nu$, and
$\chi^2_{N-1}$ is the reduced $\chi^2$ of a fit of a constant to the
data, $(N-1)\chi^2_{N-1}=\sum_j(v_j-\langle
v\rangle)^2/\sigma_j^2$. Here, we extend the original Lomb-Scargle
periodogram by allowing the mean to float at each frequency (Walker et
al.~1995; Nelson \& Angel 1998; Cumming, Marcy, \& Butler 1999), rather than subtracting the mean of the data $\langle v\rangle$ prior to the fit. The periodogram power measures
the improvement of $\chi^2$ when the sinusoid is included in the fit,
in a similar way to a classical F-test (see Cumming 2004 for a recent
discussion).

A typical Doppler measurement error is $3$--$5\ \mathrm{m/s}$,
although the precision of Doppler surveys continues to improve towards
$\sim 1\ \mathrm{m/s}$ (Mayor et al.~2003; Butler et
al.~2004). However, at these high precisions, stellar "jitter" becomes
the limiting factor. Jitter is observed at the few m/s level,
depending on stellar properties such as age, rotation, and level of
magnetic activity (Saar et al.~1998; Santos et al.~2000). For simplicity, we
add Gaussian noise with amplitude $\sigma=1\ {\rm m/s}$ throughout
this paper, which represents both Doppler errors and stellar
jitter. The mass which can be detected scales proportional to
$\sigma$, so that our results can be rescaled to the appropriate value
of $\sigma$. For example, a $5\ M_\oplus$ planet with $\sigma=5\ {\rm
m/s}$ has the same detectability as a $1\ M_\oplus$ planet with
$\sigma=1\ {\rm m/s}$. We have checked that this scaling applies for
the two planet case, when the mass of the hot jupiter is much greater
than the hot earth.

\begin{figure*}
\plottwo{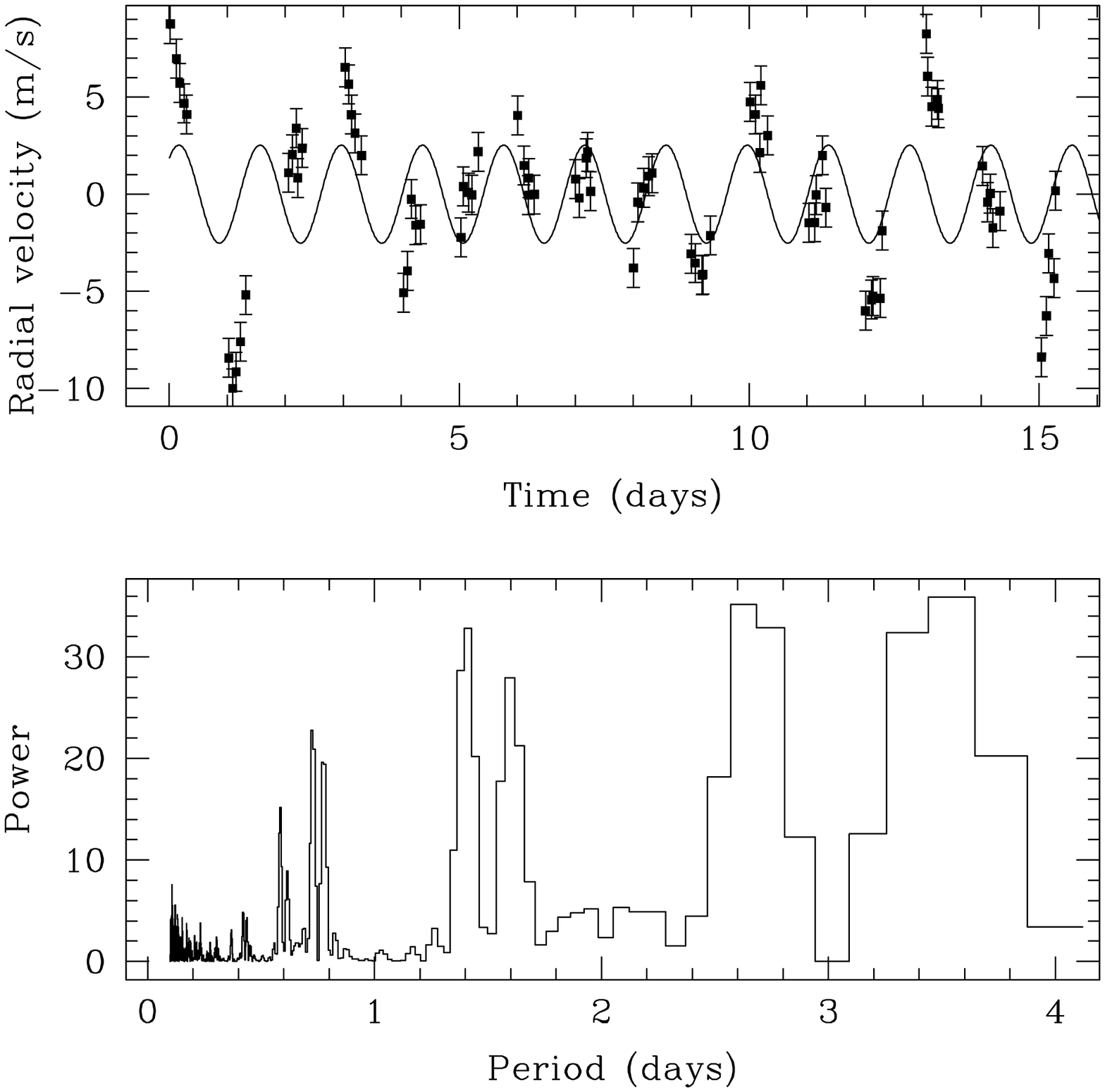}{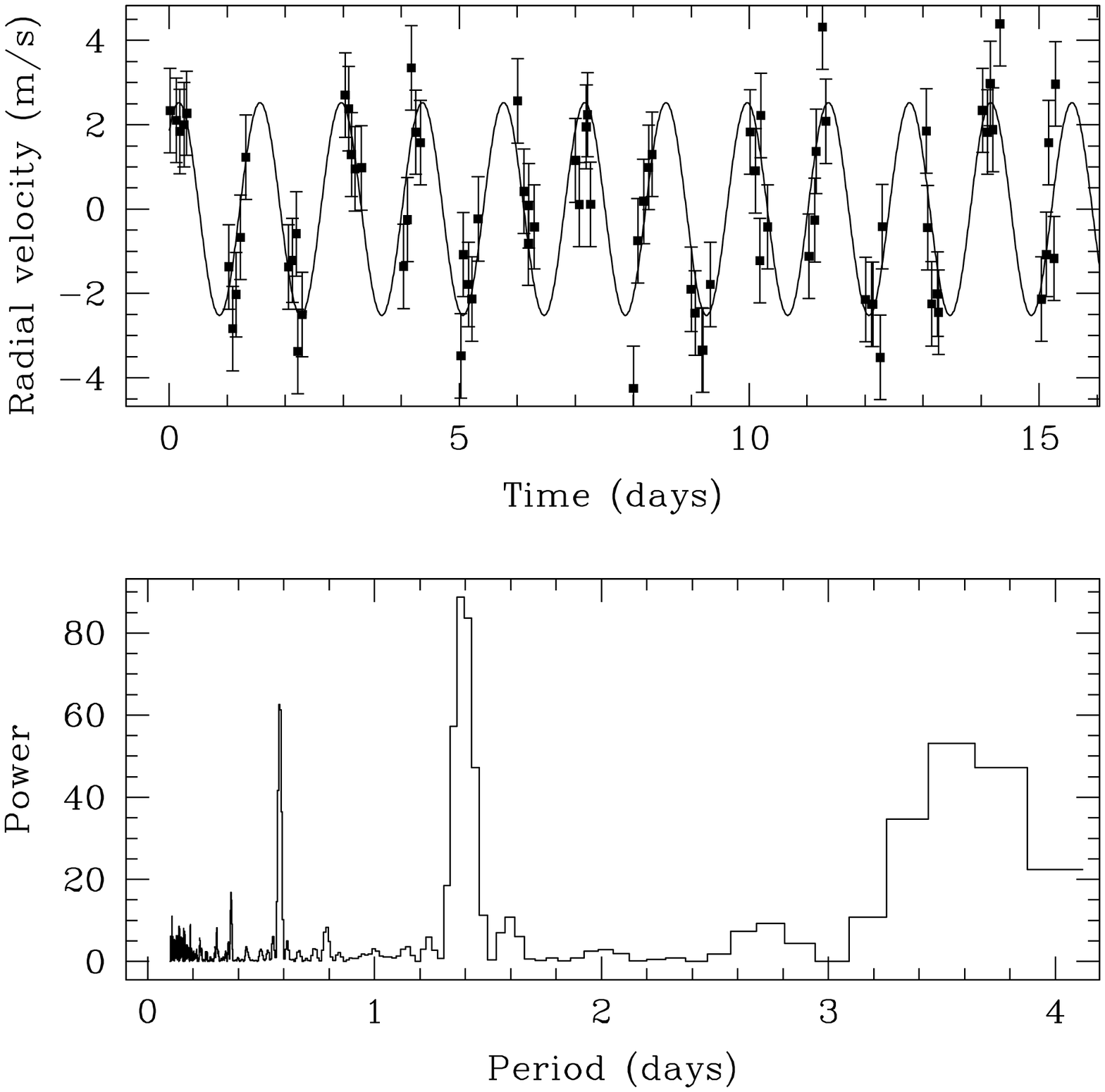}
\caption{The velocity residuals and periodogram after subtraction of
the hot jupiter for two different estimates of the hot jupiter's
orbital frequency. In the upper panel, we use the frequency estimate
from the periodogram evaluated with frequency spacing $1/4T$. In this
case, the poor subtraction leads to a large scatter in the
residuals. In the lower panel, we determine an accurate estimate of
the frequency by finding the peak of $z(\omega)$. The velocity
residuals are then dominated by the $5\ M_\oplus$
companion.\label{fig:w_brent}}
\end{figure*}

Our search procedure is to find the best-fit orbit for the hot
jupiter, subtract it from the data, and search the residuals for
periodicities. This is a much simpler approach than simultaneous
fitting of both sinusoids, and is valid as long as the parameters from
the two fits are uncorrelated. We have checked that this is the case,
and find that only for sampling rates much greater than $1\
\mathrm{day^{-1}}$ is there some improvement by simultaneous fitting. 
An alternative possibility is that the orbital solution for the hot jupiter is already known from previous measurements, in which case this information can be used to subtract the hot jupiter's signal. Therefore, we also consider the cases where the full orbital solution or only the orbital period is known in advance.

If the hot jupiter orbital period is known in advance, we find the best fit amplitude and phase at that period. If no information is available, we search for the hot jupiter by evaluating $z(\omega)$ for periods between 2 and 4 days on a grid with frequency spacing $1/4T$, where $T$ is the duration of the observations. To adequately subtract the hot jupiter's signal, it is important to accurately determine its orbital frequency.  Therefore the frequency with the largest periodogram power is used as the starting value for a more accurate search for the maximum of $z(\omega)$. Once the hot jupiter's orbit has been subtracted, we calculate $z(\omega)$ for the residuals, and find the maximum periodogram power $z_\mathrm{max}$ near the 2:1 resonance. We do not include a constant term in the fit to the residuals, giving $\nu=N-2$. Figure \ref{fig:periodogram} shows an example of the velocity residuals and the periodograms before and after subtraction. The peak at $\approx 1.5$ days in the initial periodogram is due to aliasing of the hot jupiter's frequency. After subtraction, the signal due to a $3\ M_\Earth$ planet at $1.4$ days shows clearly as a peak in $z$. 

The importance of accurately determining the hot jupiter's orbital
period is illustrated by Figure \ref{fig:w_brent}, which shows two
examples of the velocity residuals after subtraction of the hot
jupiter signal. In the first case (upper panels), we take the hot
jupiter period returned by the periodogram (evaluated with a frequency
spacing $1/4T$); in the second case (lower panels), we find a more
accurate period estimate by searching for the peak in $z(\omega)$. In
the first case, the velocity residuals are dominated by the
inadequately-subtracted hot jupiter signal rather than the hot
earth. How accurately must the hot jupiter period be known? If the
orbital frequency is known to an accuracy $\delta\omega$, the
amplitude of the residual part of the hot jupiter's signal is $\Delta
v\approx A_1\delta\omega T$, where $A_1$ is the hot jupiter
amplitude. The accuracy to which the hot jupiter frequency may be
determined is $\delta \omega\approx (2\pi/T)(\sigma/\sqrt{N}A_1)$
(e.g. Bretthorst 1988), so that
\begin{equation}\label{eq:deltav}
\Delta v\approx \frac{2\pi\sigma}{\sqrt{N}},
\end{equation}
independent of $T$ and $A_1$. For large $N$, $\Delta v\ll\sigma$, so
that the residual part of the hot jupiter signal has no effect on
detectability. However, for $N\lesssim (2\pi)^2$, the residuals from
the subtraction contribute a significant additional source of velocity
variability. Even when the hot jupiter period is known in advance, there is an uncertainty in the fitted amplitude $A_1$ of $\approx \sigma/\sqrt{N}$, and phase of $\approx (2/N)^{1/2}(\sigma/K)$, giving an additional velocity scatter for low $N$. Therefore for small $N$ we expect detection of a hot earth to be more difficult when the full orbital solution for the hot jupiter is not specified in advance.

\subsection{Calculation of Detection Threshold and Detection Probabilities}

The significance of the maximum observed periodogram power
$z_\mathrm{max}$ depends on how often an equally good or better fit
would occur purely due to a noise fluctuation. We determine this with
Monte Carlo simulations. We generate data sets with a hot jupiter plus noise, i.e.~without a hot earth, and search for a second companion. The 99\% detection threshold $z_d$ is determined as the value
of $z$ which is exceeded in only $1$\% of trials, or alternatively for
which there is a 1\% false alarm probability $F$. The detection
threshold is indicated in Figure \ref{fig:periodogram} by the dashed
line.

\begin{figure}
\plotone{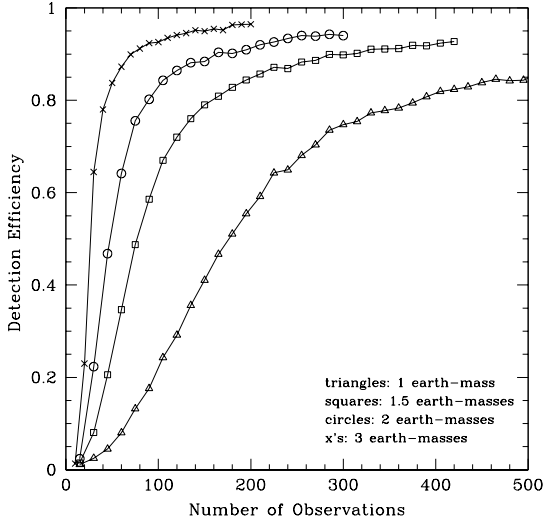}
\caption{Detection efficiency for hot earths with masses of 1, 1.5, 2,
and 3 $M_\oplus$ and period 1.4 days, with $\sigma=1\ \mathrm{m\
s^{-1}}$. We assume 1 randomly-timed observation on successive 8 hour
nights. We use 10,000 trials to evaluate the detection
probability. The hot Jupiter has a mass of 1.0 Jupiter mass, and a
period of 3.0 days.\label{fig:dp_me}}
\end{figure}

\begin{figure}
\plotone{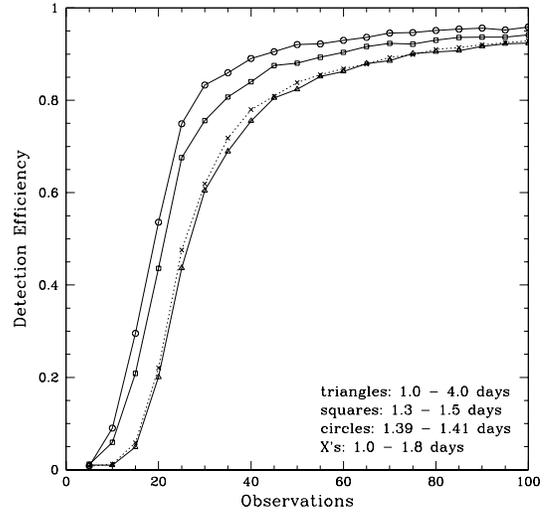}
\caption{The effect of the range in orbital period searched on the
detection probability. We take $\sigma = 1 \mathrm{m s^{-1}}$ and
$M_P=3\ M_\earth$. Triangles represent orbital periods ranging from 1
-- 4 days, squares 1.3 -- 1.5 days, and circles 1.39 -- 1.41 days.
The dotted line is for 1 -- 1.8 days, which is used elsewhere in this
paper. As a larger frequency range is searched, $N_i$ increases giving
a larger detection threshold and lower detection
probability. \label{fig:dp_frange}}
\end{figure}

\begin{figure}
\plotone{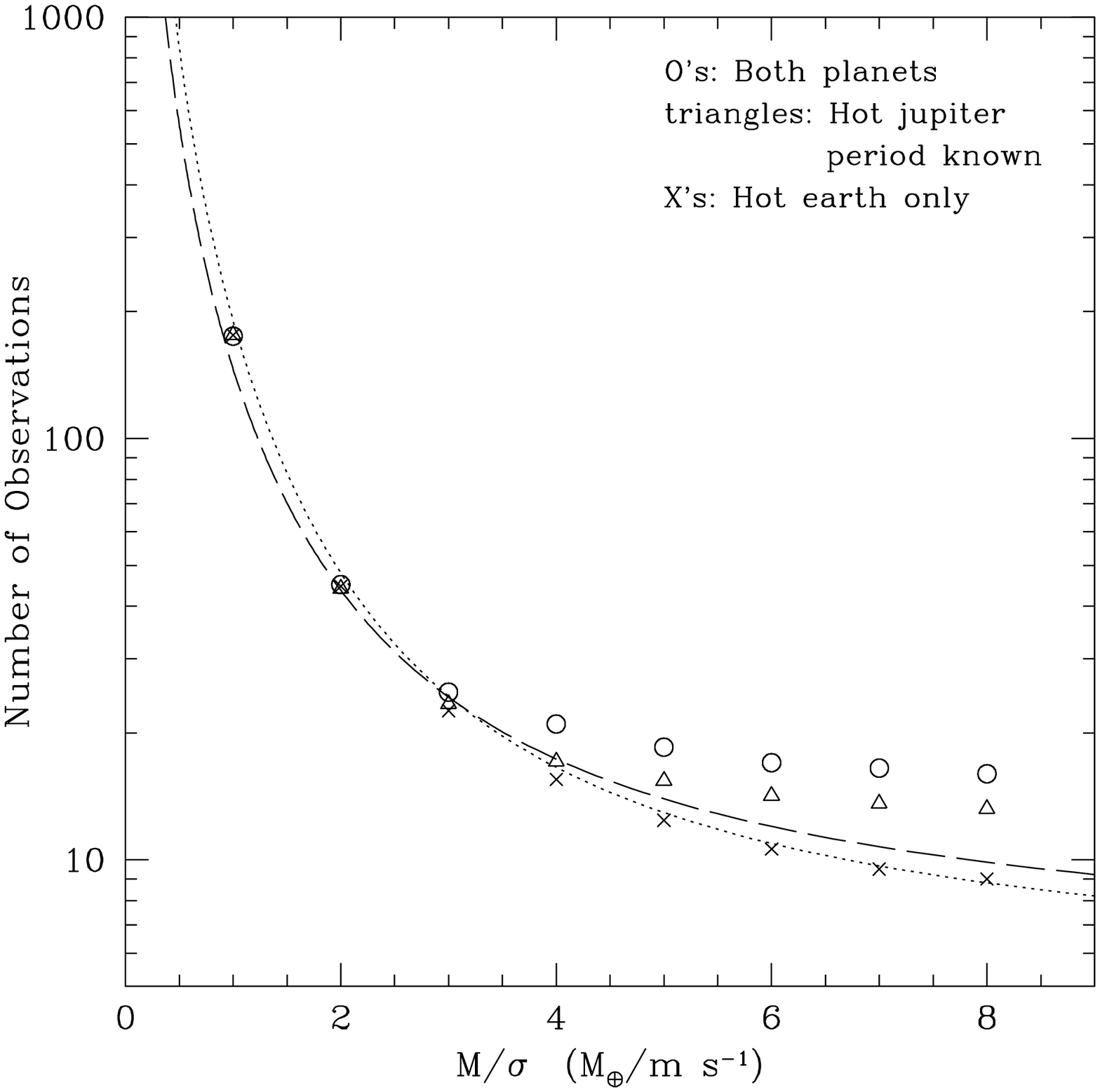}
\caption{The number of observations needed to detect a planet 50\% of the time as a function of the mass $M_P$ scaled by the error $\sigma$. Observations are taken at a random time during each consecutive night. The crosses are for data sets containing only a hot Earth, the triangles for the case where the hot jupiter period is known in advance, and the circles for a full search for both orbital periods. The curves are the analytic results. The dotted curve occurs if the number of independent frequencies, $N_i$, is $N_i = T \Delta f$, where $T$ is the duration and $\Delta f$ is the frequency range used to search for the hot earth.  For the dashed curve, $N_i = 10$.\label{fig:mass}}
\end{figure}

We adopt a similar Monte Carlo approach to determine the detection
probability. We generate a large number of data sets for each choice
of $M_p$ and $N$, and calculate the fraction of trials for which the
hot earth is detected. A random distribution of inclinations is
included. Figure \ref{fig:dp_me} shows the detection efficiency as a
function of number of observations $N$ for $M_P=1$, 1.5, 2, and $3\
M_\Earth$. We simulate 1 observation per 8 hour night,
and take observations for successive nights. In these simulations, we
search for the hot earth orbital period between $1$ and $1.8$
days. Figure \ref{fig:dp_frange} shows the effect of changing this
frequency range. If the frequency of the hot earth is specified in
advance, the detectability is increased since the number of
``independent frequencies'' in the search is less, giving a smaller
chance of a false alarm due to a noise fluctuation. This is the
well-known ``bandwidth penalty'' (e.g.~Vaughan et al.~1994).

In Figure \ref{fig:mass}, we show the number of observations needed to detect a given mass 50\% of the time. The crosses are for data sets containing only a hot Earth, or equivalently, for the case when the hot jupiter orbit is completely known in advance. The triangles are for the case with a hot jupiter whose orbital period is known in advance, and the circles for the case with a hot jupiter, but with a full search for both orbital periods. For $N\approx 10$--$20$, planets with $M\gtrsim 4\,(\sigma/1\ {\rm m\ s^{-1}})\ M_\oplus$ can be detected. However, detection of a planet with  $M\sim 1\,(\sigma/1\ {\rm m\ s^{-1}})\ M_\oplus$ requires $N\approx 200$. For $N\lesssim 20$, detection of the hot earth after fitting for the hot jupiter's signal is harder than detection of the hot earth alone or with the hot jupiter orbit fully specified in advance (compare the crosses and circles in Fig~\ref{fig:mass}). This is likely due to the additional velocity scatter at low $N$ from inadequate subtraction of the hot jupiter's signal ($\Delta v\gtrsim \sigma$ in eq.~[\ref{eq:deltav}]).

\subsection{Analytic estimates}

In this section, we derive an analytic estimate for the detection
threshold, following the approach of Cumming (2004),
who discusses the detectability of single planets with radial
velocities. First, we determine $z_d$ semi-analytically for Gaussian
noise. The cumulative distribution of $z(\omega)$ for a single
frequency is (e.g.~Cumming et al.~1999)
\begin{equation}\label{eq:prob}
{\rm Prob}(z>z_0)=\left(1+{2z_0\over\nu}\right)^{-\nu/2},
\end{equation}
or for large $N$,
\begin{equation}
{\rm Prob}(z>z_0)\approx\exp(-z_0),\label{eq:prob2}.
\end{equation}
For a given false alarm probability $F$, the detection threshold is given by
\begin{equation}\label{eq:FA}
F=1-\left[1-{\rm Prob}(z>z_d)\right]^{N_i},
\end{equation}
where $N_i$ is the number of ``independent frequencies'' searched. For $F\ll 1$,
\begin{equation}
F\approx N_i\ {\rm Prob}(z>z_d).
\end{equation}
For unevenly-sampled data, $N_i$ must be determined by Monte Carlo
simulations. However, since the spacing of periodogram peaks is $1/T$,
a rough estimate is $N_i\approx T\Delta f$, where $\Delta f$ is the
frequency range searched (Cumming 2004).

We next make an analytic estimate of the detection threshold. In the
presence of a signal with amplitude $K$, the average power is $z_s=\nu
K^2/4\sigma^2$ (Groth 1975; Scargle 1982; Horne \& Baliunas 1986, we
have accounted for the different normalization). For a given $N_i$ and
$F$, we find the detection threshold $z_d$ by inverting equation
(\ref{eq:FA}) using the analytic distribution (\ref{eq:prob}). Then,
setting $z_s=z_d$ gives the signal to noise ratio needed to detect the
signal
\begin{eqnarray}\label{eq:sn1}
{K\over\sqrt{2}\sigma}&=&\left[\left({N_i\over
F}\right)^{2/(N-3)}-1\right]^{1/2}\\ &\approx& \left[2\ln(N_i/F)\over
N\right]^{1/2},\label{eq:sn2}
\end{eqnarray} 
(Cumming et al.~2003; Cumming 2004) where the approximation is for large
$N$. As expected, the signal to noise ratio goes down as $1/\sqrt{N}$
for $N\gg 1$.

\begin{figure*}
\plottwo{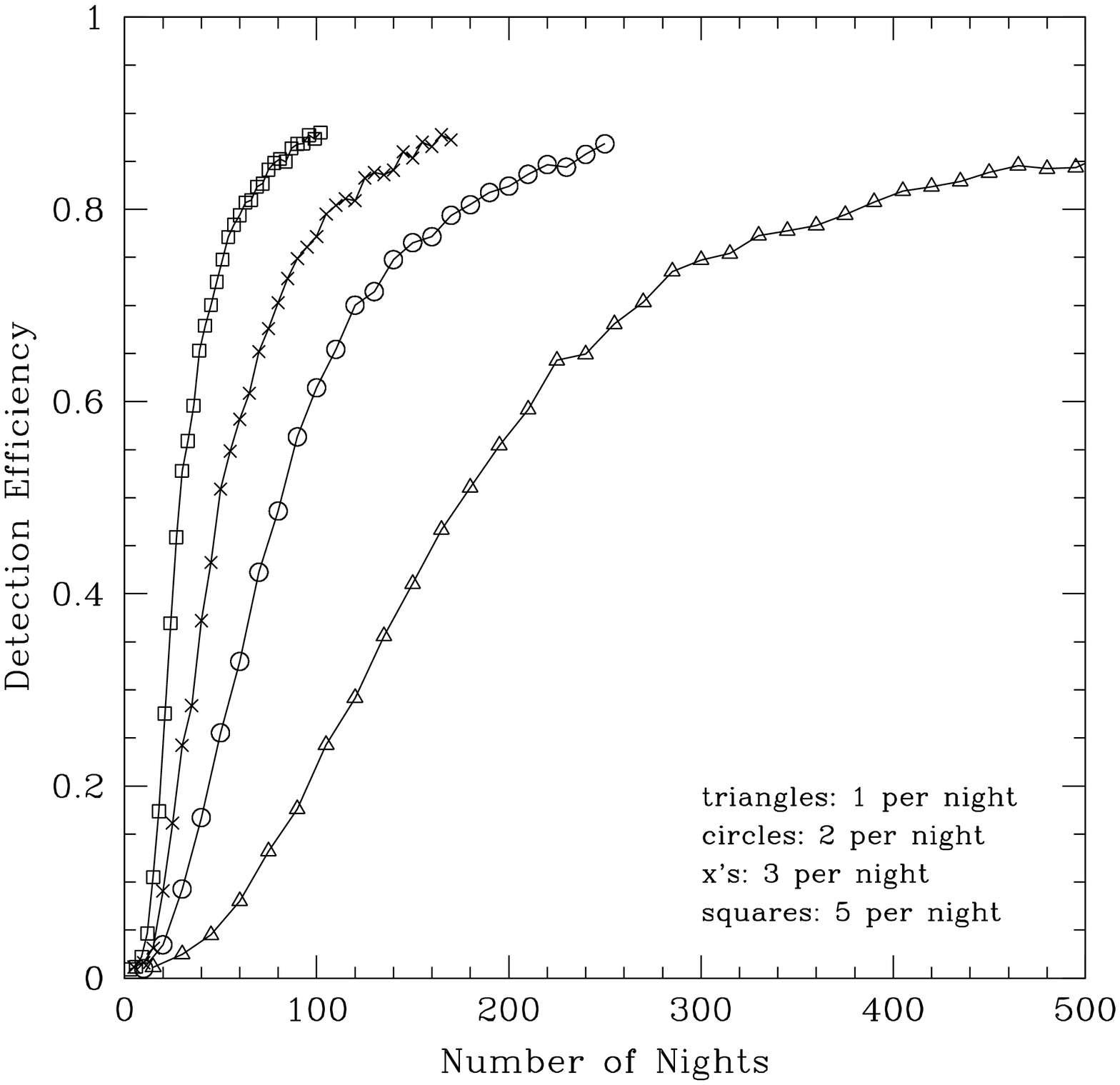}{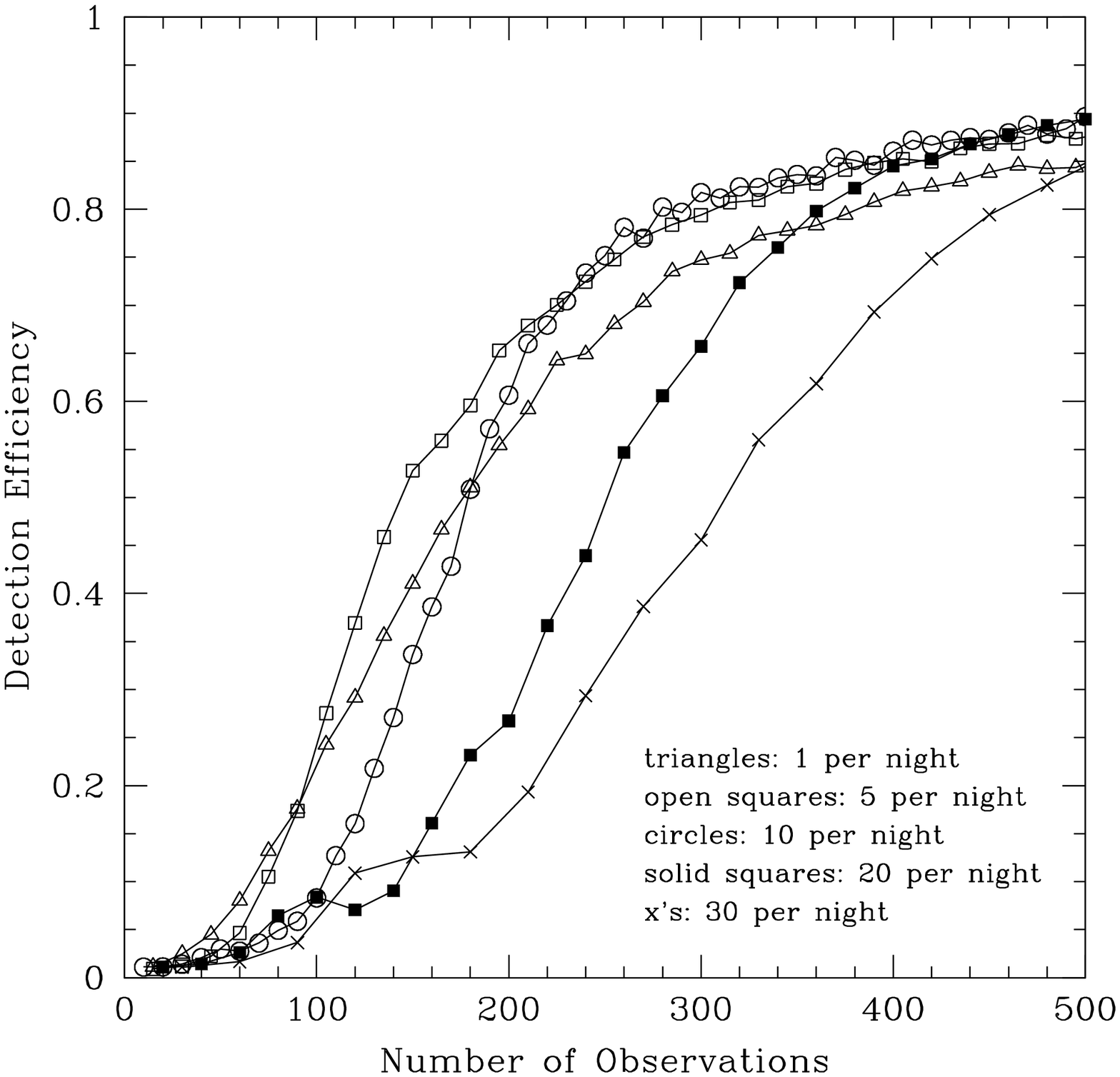}
\caption{Detection probability for a $1\ M_\earth$ planet with
different numbers of observations made on consecutive nights. We show
the detection probability against number of nights (left panel) and
number of observations (right panel). A more rapid sampling rate leads
to a detection in fewer nights, but at the cost of taking many
observations.\label{fig:dp_days}}
\end{figure*}

Combining equation (\ref{eq:K}) with equation (\ref{eq:sn1}) or
(\ref{eq:sn2}) gives the velocity amplitude and mass which can be
detected 50\% of the time. Equation (\ref{eq:sn2}) ($N\gg 1$) gives
\begin{equation}\label{eq:K50}
K_{50}={6\ {\rm m\ s^{-1}}\over \sqrt{N}}\ \left({\sigma\over{\rm m\
s^{-1}}}\right)\left({\ln [N_i/F]\over 9.2}\right)^{1/2},
\end{equation}
or
\begin{equation}\label{eq:M50}
M_{50}\approx {10\ M_\Earth\over \sqrt{N}}\left({\sigma\over
\mathrm{m\ s^{-1}}}\right)
\left({P\over\mathrm{d}}\right)^{1/3}\left({\ln [N_i/F]\over
9.2}\right)^{1/2} \left({M_\star\over M_\odot}\right)^{2/3},
\end{equation}
where we have also used the fact that the mean value of $\sin i$ is
$\pi/4$, and we take $N_i=100$ and $F=0.01$.

Figure \ref{fig:mass} compares the analytic results for arbitrary $N$
(eq.~[\ref{eq:sn1}]) with our numerical simulations. The dashed and dotted curves show the analytic models for two
different estimates of $N_i$. The agreement is excellent for a single
planet. For the two planet case, equation (\ref{eq:M50}) applies for
$N\gtrsim 20$, but underestimates $M_{50}$ for $N\lesssim 20$. For
such low $N$, we expect $\Delta v\gtrsim \sigma$
(eq.~[\ref{eq:deltav}]), making detection of the hot earth more
difficult than in the single planet case.

\subsection{Observing Strategy}

We now address to what extent the spacing of the observation times
affects the detectability of the hot earth. First, we consider
observations made on successive nights, but with different numbers of
observations per night. Figure \ref{fig:dp_days} shows the detection
probability as a function of number of nights, and as a function of
number of observations. For a fixed number of observations, increasing
the sampling rate reduces the duration of the data, leading to less
frequency resolution and therefore smaller $N_i$. We therefore expect
better detectability for a faster sampling rate at a fixed number of
observations. This is initially the case, as can be seen by comparing
the curves for 1 per night and 5 per night in Figure
\ref{fig:dp_days}. However, for very rapid sampling, a large number of
observations must be made before a complete hot jupiter orbit is
sampled, therefore allowing the orbital parameters to be determined
accurately enough for a good subtraction. This is likely the reason
for the decrease in detectability on going to more than 5 observations
per night in Figure \ref{fig:dp_days}.  Although observing more
rapidly always leads to detection in a fewer number of nights (left
panel of Fig.~\ref{fig:dp_days}), the number of observations required
eventiually becomes very large (right panel of
Fig.~\ref{fig:dp_days}). Therefore there is an optimum observing rate
(roughly a few per night for this case, but it depends slightly on the
hot earth mass).

Although observing on successive nights would be possible with a
dedicated telescope, current radial velocity surveys are limited by
telescope scheduling. Therefore, we have also simulated more realistic observation times for
current surveys, by observing for three nights each month, with the
seperation between observing runs randomly chosen between 20 and 40
days. We find that if the hot jupiter orbital parameters are not known in advance, the detectability can be affected by the observing scheme, because of aliasing of the hot jupiter frequency leading to subtraction of an incorrect orbit. However, in the practical case that the hot jupiter period is known from previous observations, the hot jupiter orbit can be adequately subtracted, and the effect of the observing strategy is small.

\begin{deluxetable*}{llllllllll}
\tablewidth{0pt}
 \tablecaption{Summary of radial velocity data for
hot jupiters and estimated upper limits for hot earths}
\tablehead{
 \colhead{Star name} & \colhead{References} &
\colhead{$P_J$\tablenotemark{a}} & \colhead{rms\tablenotemark{b}} &
\colhead{$N$} & \colhead{$T$} & \colhead{$N_i$\tablenotemark{c}} &
\colhead{$M_\star$} &\colhead{$K_{99}$\tablenotemark{d}} &
\colhead{$M_{99}$\tablenotemark{e}} \\ \colhead{} & \colhead{} &
\colhead{(d)} & \colhead{$({\rm m\ s^{-1}})$} & \colhead{} &
\colhead{(d)} & \colhead{} & \colhead{$(M_\odot)$} &\colhead{$({\rm m\
s^{-1}})$} & \colhead{$(M_\Earth)$}}
 \startdata HD 73256 & 1 & 2.55
& 15 & 40 & 80 & 4 & 1.05 & 22 & 58\\ HD 83443 & 2,3 & 2.99 & 3.8 & 36
& 1150 & 38 & 0.79 & 7.3 & 15\\ HD 46375 & 4 & 3.02 & 2.6 & 24 & 516 &
21 & 1.0 & 6.5 & 15\\ HD179949 & 5 & 3.09 & 10 & 23 & 735 & 30 & 1.24
& 27 & 74\\ HD 187123 & 6 & 3.10 & 7.6 & 20 & 250 & 10 & 1.0 & 20 &
48\\ $\tau$ Boo & 7 & 3.31 & 16.5 & 58 & 4400 & 166 & 1.3 & 26 & 74 \\
BD-103166 & 8 & 3.49 & 8.1 & 17 & 410 & 15 & 1.1 & 26 & 69\\ HD 75289
& 9 & 3.51 & 7.5 & 88 & 330 & 12 & 1.15 & 7.7 & 21\\ HD 76700 & 10 &
3.97 & 6.2 & 24 & 1243 & 39 & 1.0 & 16 & 42\\ \enddata
\tablenotetext{a}{Orbital period of the hot jupiter.}
\tablenotetext{b}{rms of the residuals to the orbital solution.}
\tablenotetext{c}{Estimated number of independent frequencies for a
search from $2$--$2.4$ times the hot jupiter orbital frequency,
$N_i\approx 0.1 T/P_J$.}  \tablenotetext{d}{Estimated 99\% upper
limit, using equation (\ref{eq:sn1}) with $K_{99}\approx 1.7 K_{50}$.}
\tablenotetext{e}{99\% upper limit on the mass of a hot earth near the
2:1 resonance with the hot jupiter.}
\tablenotetext{f}{References.---(1) Udry et al.~2003; (2) Mayor et
al.~2002; (3) Butler et al.~2002; (4) Marcy et al.~2000; (5) Tinney et
al.~2001; (6) Butler et al.~1998; (7) Fischer et al.~2001; (8) Butler
et al.~2000; (9) Udry et al.~2000; (10) Tinney et al.~2003}
\end{deluxetable*}

\section{Summary and Discussion}

We have calculated the detectability of low mass planets in radial
velocity surveys, and in particular a ``hot earth'' companion to a hot jupiter. Detection of such a companion would give important clues to the ordering of the planet formation process, and the ubiquity of terrestrial mass cores around solar type stars. The velocity amplitude and mass required for a 50\% detection rate are given by equations (\ref{eq:K50}) and (\ref{eq:M50}), and shown in Figure
\ref{fig:mass}. For $N\approx 20$, masses greater than $4\ M_\oplus\
(\sigma/{\rm m\ s^{-1}})(P/{\rm d})^{1/3}(M_\star/M_\odot)^{2/3}$ have
50\% detectability or better. 

\begin{figure*}
\epsscale{1.0}
\plottwo{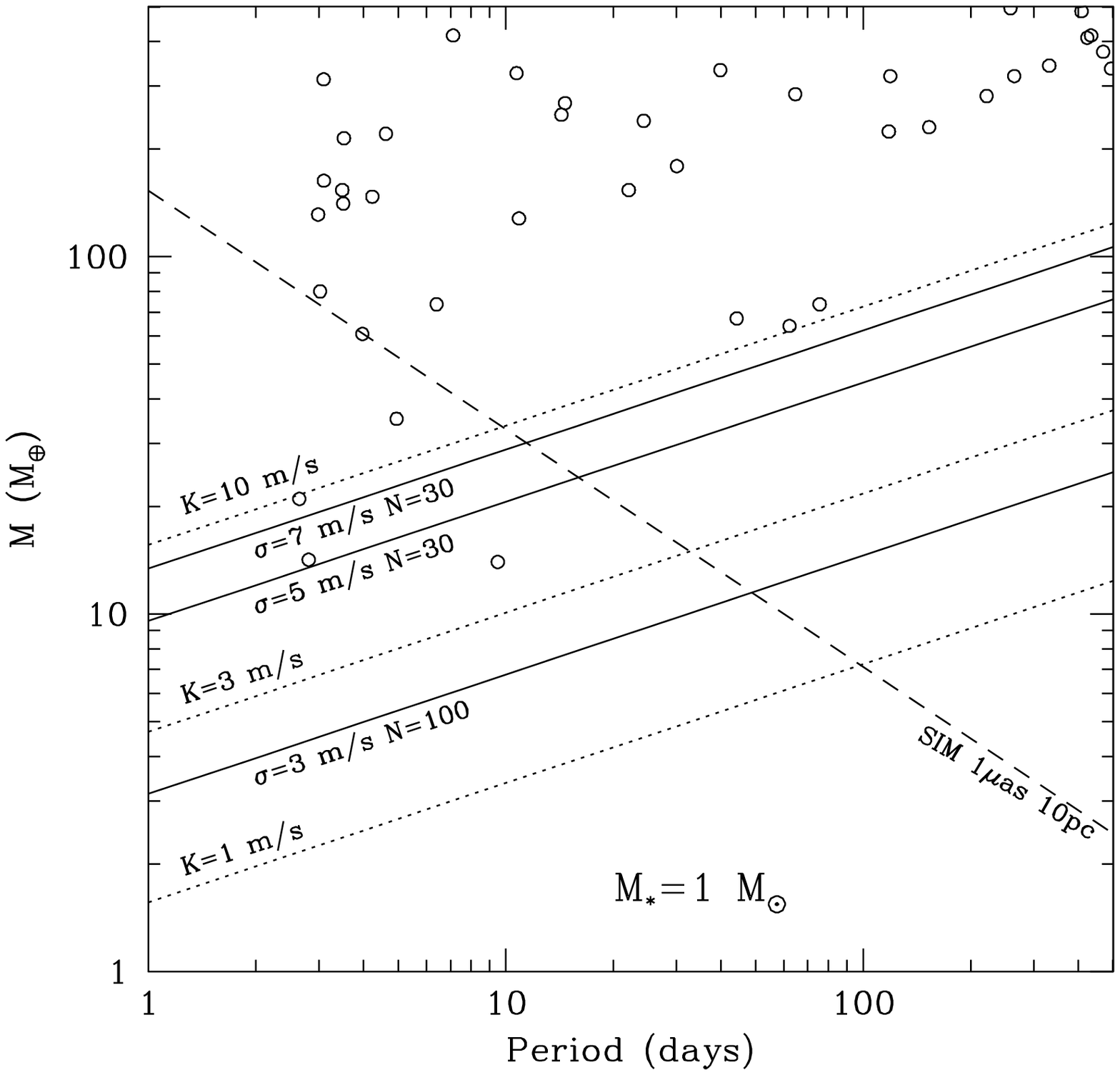}{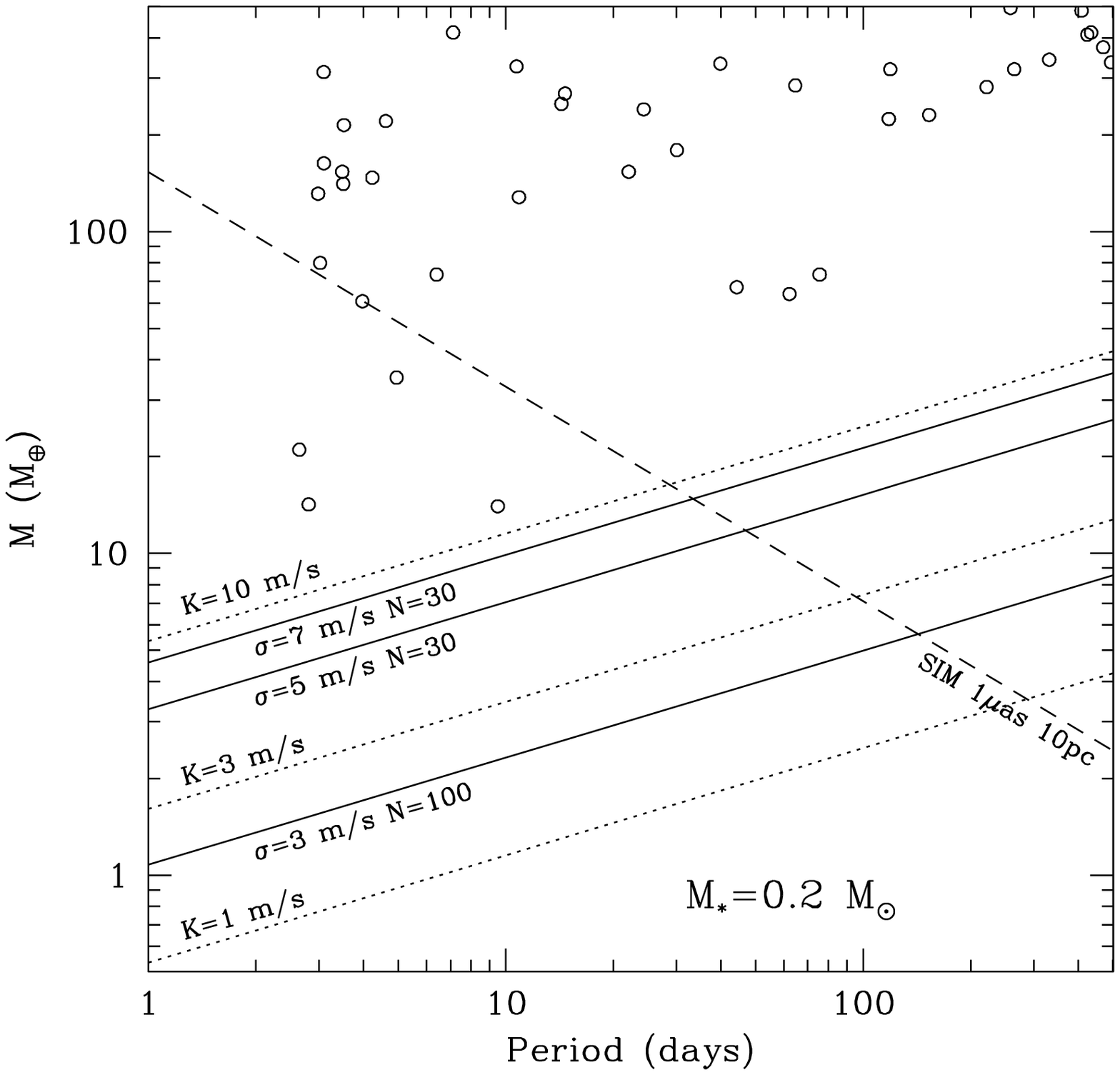}
\caption{Summary of detection thresholds in the mass-period plane. The
dashed lines show 50\% detection thresholds for different $N$ and
$\sigma$, for $M_\star=1\ M_\odot$ (left panel) and $M_\star=0.2\
M_\odot$ (right panel). It is assumed that the duration of the
observations is longer than the orbital period. The circles show
currently detected planets ($M\sin i$) from exoplanets.org, and including the 3 recently announced Neptune-mass candidates. The dotted
lines show velocity amplitudes of $K=1,3,$ and $10\ \mathrm{m/s}$
($\sin i=1$), and the dashed line shows an approximate detection
threshold for SIM (assuming $1\mu\mathrm{as}$ sensitivity and a $10\
\mathrm{pc}$ distance; see Ford \& Tremaine 2003).\label{fig:summary}}
\end{figure*}

Our results apply to a low mass planet that is near the mean motion resonance
with a hot jupiter, or that is isolated. In Figure
\ref{fig:summary}, we show the detection limits of equation
(\ref{eq:M50}) compared to the present distribution of known
exoplanets\footnote{Taken from {http://www.exoplanets.org/}.}. The left panel is for
$M_\star=1\ M_\odot$, the right panel for $M_\star=0.2\
M_\odot$. Lower mass stars have a greater velocity amplitude for a
given planet mass, so that M dwarfs are the most promising to search
for terrestrial mass bodies. Close orbits are particularly interesting
for this case, since they lie within the habitable zone for M dwarfs
(orbital periods of a few to tens of days; Kasting, Whitmire, \&
Reynolds 1993; Joshi, Haberle, \& Reynolds 1997).

Recently, three Neptune-mass planets have been discovered in close orbits around GJ~436 (Butler et al.~2004; $M\sin i=21\ M_\Earth$, $P=2.644$d, $N=42$, rms=$5.3\ {\rm m\ s^{-1}}$), $\rho$ Cnc (McArthur et al.~2004; $M\sin i=14.2\ M_\Earth$, $P=2.808$d, $N=119$, rms=$5.4\ {\rm m\ s^{-1}}$), and $\mu$ Ara (Santos et al.~2004; $M\sin i=14\ M_\Earth$, $P=9.5$d, $N=24$, rms=$0.9\ {\rm m\ s^{-1}}$). These can be seen as the lowest mass planets in Figure \ref{fig:summary}. Only one of these stars, GJ~436, is an M dwarf. All of these detections lie above the detection limits in Figure \ref{fig:summary}.

The lowest mass planet detected prior to these recent announcements was 
HD 49674b (Butler et al.~2002) with $M\sin i=0.12\ M_J$ (or $\approx
40\ M_\oplus$), $K=13\ {\rm m\ s^{-1}}$, and $P=5\ {\rm days}$. The
velocity curve presented by Butler et al.~(2002) has $N=24$, and the
residual rms is $\approx 5\ {\rm m\ s^{-1}}$. With these values of $N$
and $\sigma$, the thresholds we derive here imply that planets with
half the mass of HD~49674b ($\approx 10$--$20\ M_\oplus$ depending on
orbital period) should be detectable, consistent with the recent detections. Endl et al.~(2003) report 22
velocity measurements of the M dwarf Proxima Cen with Doppler errors
of $2.5\ {\rm m\ s^{-1}}$. They calculate an upper limit of $4$--$6\
M_\earth$ for planets in the habitable zone (estimated to be $4$--$14$
days), in good agreement with Figure \ref{fig:summary}.

A summary of published data for hot jupiters with orbital periods
$P_J<4$ days is shown in Table 1. To estimate an upper limit on the
velocity amplitude of a hot earth, we first use the duration and
number of observations to estimate $N_i$ for a search between orbital
periods $P_J/2.4$ and $P_J/2$. We then write the 99\% upper limit as
$K_{99}=1.7 K_{50}$, where $K_{50}$ is given by equation
(\ref{eq:sn1}), but with the rms of the residuals to the best-fitting
hot jupiter orbit substituted for $\sigma$. Since we adopt the full
rms of the residuals as the noise level, these are conservative upper
limits. The values of $K_{99}$ range from $7$--$30\ {\rm m\ s^{-1}}$,
corresponding to $\approx 15$--$80\ M_\earth$ for companions close to
2:1 resonance with the hot jupiter. HD 83443 and HD 46375 have upper
limits of $15\ M_\earth$ because of very precise measurements giving a
low rms (4 and $3\ {\rm m\ s^{-1}}$ respectively).

These detections and upper limits imply that the current threshold for detecting planets in close orbits is $\gtrsim 10$--$20\ M_\Earth$, with planets close to this threshold just starting to be discovered. With more observations, and improvements in measurement precision, this threshold can be pushed lower. In Figure \ref{fig:summary}, we show the detectable mass for $N=100$ and $\sigma=3\ {\rm m\ s^{-1}}$,
which reaches a few Earth masses for orbits $\sim 1$ day for a solar
mass star, or $1\ M_\earth$ for a $0.2\ M_\odot$ M dwarf. However, a better understanding of stellar jitter will be crucial for
convincing detections of small amplitude planets. Detection of extrasolar planets with radial velocities becomes more
and more difficult as $K$ approaches $\sigma$, because alternative
explanations for the detected signal must be excluded (see the
discussion in Cumming et al. 1999 for example). These include systematic errors in
the measurements, as well as stellar jitter. Both may be periodic, with
timescales associated with seasonal variations, or stellar properties
such as rotation, convective motions, or the appearance and
disappearance of magnetic features. Saar \& Fischer (2000), Paulson et
al.~(2002), and Saar (2003) discuss attempts to correct for stellar
jitter using simultaneous measurements of activity indicators or spectral line shapes, but
such studies are just beginning. Very precise Doppler measurements
with errors at the $1\ {\rm m\ s^{-1}}$ level will help to understand
the stellar jitter and systematic errors. 

Finally, we have assumed that both planets are in circular orbits in this paper.
For orbital periods as short as a few days, the timescale for tidal
circularization is expected to be much shorter than the age of the system.
However, additional planets in the system, at longer orbital periods and
so far undetected, may excite the eccentricity of the hot jupiter. Tidal
damping of an eccentricity induced in this way has been suggested as the
reason for the inflated radius of the transiting planet HD~209458b
(Bodenheimer, Lin, \& Mardling 2001). The required eccentricity is
comparable to current measurement uncertainties, typically $\delta e\sim
0.03$. Fitting a circular orbit to a Keplerian orbit with eccentricity $e$
gives a residual scatter $\approx eA_1$, where $A_1$ is the amplitude.
For a hot jupiter with $A_1\approx 100\ {\rm m\ s^{-1}}$ the
scatter in the residuals is comparable to the signal from a hot earth.
However, since the Fourier components of a Keplerian orbit are at the
orbital period and its harmonics, whereas the hot earth is expected to lie
outside the 2:1 resonance, it should be possible to distinguish
between these two possibilities, although simultaneous fitting of the hot
jupiter and earth orbits may be required. Therefore we do not expect a
significant reduction in sensitivity over the detection thresholds
calculated in this paper. This issue is closely related to the
important question of determining the uncertainty in orbital parameters (Ford 2003),
particularly the eccentricity of hot jupiters, and deserves further study.

\acknowledgements 
We acknowledge support from NSF through grant AST-9987417 and NASA through grant NAG5-13177. AC is supported by NASA through Hubble Fellowship
grant HF-01138 awarded by the Space Telescope Science Institute, which is operated by the Association of Universities for Research in Astronomy, Inc., for NASA, under contract NAS 5-26555.

\end{document}